\documentclass[aps,superscriptaddress,amsfonts,twocolumn,showpacs]{revtex4-2}
\usepackage[colorlinks=true]{hyperref}
\usepackage{mathtools}
\newcommand{\Kt}{{\rm K}}
\newcommand{\PDF}{{\mathcal{P}}}
\newcommand{\ft}{\widetilde{f}}

\begin{document}

\title{Soliton gas resolution and  statistics of  random wave fields in semiclassical integrable turbulence}

\author{T. Congy}
\email{thibault.congy@northumbria.ac.uk}
\author{G. A. El}
\affiliation{School of Engineering,  Physics and Mathematics, Northumbria University, Newcastle upon Tyne NE1 8ST, United Kingdom}

\begin{abstract}

We develop a general analytical framework for determining the probability distribution of random nonlinear wave fields governed by the focusing nonlinear Schr\"odinger equation (fNLSE)  in  regimes where typical realizations are dominated by solitons.  We formulate the soliton gas resolution conjecture for the long-time evolution of slowly varying (“semiclassical”) random initial states and implement a stochastic analogue of the inverse scattering transform by establishing a relationship between the spectral density of states of the underlying bound state soliton gas and the probability density function (PDF) of the  intensity of the resulting turbulent wave field.
The derived  explicit integral representation for the PDF is shown to be in excellent agreement with direct numerical simulations across several representative examples of fNLSE integrable turbulence. {The analytical results of the paper are exact; they do not involve any closure assumptions, in contrast to classical turbulence theories. The results have broad applicability to physical systems including water waves, nonlinear optics, and superfluids, where integrable equations serve as important asymptotic models.}

\end{abstract}

\maketitle

{Integrable turbulence (IT) and soliton gases have recently emerged as a topical research area, driven by both the depth and richness of the underlying mathematical structures and the recognition of the ubiquity of  these phenomena across a broad range of physical systems — from surface and internal water waves \cite{costa_soliton_2014, Redor:21,    lee_field_2026, leduque_deep_2025, michel_emergence_2020}  to light propagation in optical fibers \cite{walczak_optical_2015, suret_single-shot_2016, kraych_statistical_2019, fache_perturbed_2025}, turbulence in collisionless plasmas \cite{meiss_drift-wave_1982}, and quantum dynamics in ultracold atomic gases \cite{mossman_observation_2024, ballu_soliton_2026}. See the review articles \cite{el_soliton_2021, suret_soliton_2024} for the recent theoretical and experimental developments.}

 {IT, introduced by V. Zakharov \cite{zakharov_turbulence_2009} as a statistical theory of random nonlinear wave fields governed by integrable nonlinear dispersive equations, provides a critical leading-order approximation to strong wave turbulence in various applications, where integrable equations such as the Korteweg-de Vries (KdV) or nonlinear Schr\"odinger (NLS) equation are important physical models. As such, the IT framework holds the
compelling prospect of yielding solvable, physical turbulent models. In contrast, classical fluid turbulence is hampered by the closure problem, in which the evolution equations for statistical moments of the fluid state involve successively higher-order moments, yielding an infinite hierarchy that must be closed by ad hoc approximation \cite{pope_turbulent_2000}.
In addition to its intrinsic theoretical significance, the IT framework is of particular relevance to the prediction and characterization of extreme events such as rogue waves in the ocean \cite{onorato_freak_2001, kharif2008Rogue} or in optical media \cite{akhmediev_recent_2013, dudley2019rogue}.}

A canonical example of the IT occurrence is  the development  of the noise induced (spontaneous) modulational instability of the plane wave solution of the focusing NLS equation (fNLSE)
\begin{equation}\label{fNLSE}
i\psi_t+\psi_{xx}+2|\psi|^2\psi=0, \quad \psi \in \mathbb{C}
\end{equation} 
studied numerically in \cite{agafontsev2015integrable}, where it was shown that such evolution results, in the long time regime, in the emergence of  statistically stationary IT,  locally  exhibiting strongly nonlinear coherent structures such as solitons and breathers but nevertheless,  somewhat paradoxically, characterized by the Gaussian single-point statistics. 

Another important example  of the  IT emergence is the long-time  evolution of the so-called partially coherent waves  typically generated in optical systems as narrowband extended signals composed of many Fourier modes with random, statistically independent phases  implying, by the Central Limit Theorem,  the Gaussian statistics for the resulting slowly varying random amplitude field. In this case, the fNLSE  evolution was numerically observed in \cite{Tikan2020Effect,agafontsev_extreme_2021} to lead to the emergence of statistically stationary IT but now the long-time equilibrium statistics is strongly non-Gaussian, exhibiting ``heavy tails''--- the signature of the presence of rogue waves \cite{Onorato2013Rogue, randoux_nonlinear_2016}.

Until recently, theoretical research on IT has been mostly limited to numerical simulations {(see, e.g., \cite{agafontsev2015integrable, agafontsev2016integrable, Tikan2021Local}) revealing a number of intriguing statistical features related to the presence of coherent structures and markedly different from the statistics of strong hydrodynamic or weak wave turbulence \cite{pope_turbulent_2000, nazarenko_wave_2011}}. 
Notable analytical progress became possible  with the development of the spectral theory of soliton gases—--large ensembles of interacting solitons characterized by random distributions of their parameters and most naturally described within the inverse scattering transform (IST) framework \cite{el_soliton_2021, suret_soliton_2024}. Within this framework, individual solitons correspond to points $\lambda_j$ of discrete spectrum of the linear Lax operator associated with the integrable PDE. Soliton gases therefore constitute a special class of IT, in which the turbulent state is formed by a macroscopic ensemble of randomly distributed solitons.

For the fNLSE  \eqref{fNLSE} the solitonic IST (Zakharov-Shabat) spectrum is given by c.c. pairs of eigenvalues $\{\lambda_j, \bar{\lambda}_j\}$,  $\lambda_j = \xi_j +  i \eta_j$, $\eta_j>0$. It was shown in \cite{bronski_semiclassical_1996, tovbis_semiclassical_2004}  that the IST spectrum of slowly varying $L^2$ potentials 
\begin{equation}\label{madelung}
\psi(x,0) = \sqrt{\rho_0(x)} \, e^{i\int u_0(x)\,dx}
\end{equation}
satisfying
\begin{equation} \label{slow_var}
\frac{\rho_0'(x)}{\rho_0(x)} = { O}(\ell^{-1}),\quad u_0(x) =  { O}(\ell^{-1}),\quad  \ell \gg 1,
\end{equation}
---is dominated by discrete eigenvalues located close to the imaginary axis,  $\lambda_j \simeq i \eta_j$, so that the resulting multisoliton ensembles represent to leading order non-propagating bound states.
This result,  obtained in the semiclassical limit of fNLSE,   motivates the {\it soliton gas resolution conjecture} for  nondecaying ergodic random potentials satisfying the slow variation condition \eqref{slow_var}. Specifically, we conjecture that the semiclassical IT developing in the long-time evolution of random potentials \eqref{madelung}, \eqref{slow_var} is accurately approximated by the bound state soliton gas.  The numerical results in \cite{gelash2019bound, agafontsev_extreme_2021} for the above two prototypical classes of IT (long-time development of the spontaneous modulational instability and the evolution of  partially coherent waves) provide strong numerical evidence in support of this conjecture.

\smallskip
In this paper, we {address one of the  outstanding problems in integrable turbulence, which has so far  been approached mostly numerically, with the exception of some partial analytical results in \cite{congy_statistics_2024, biondini_breather_2025}.} Specifically, we  analytically determine  the probability density function (PDF) $\mathcal{P}_\infty(\rho)$ of the wave field intensity $\rho= |\psi|^2$ in the fully developed, statistically stationary   IT 
emerging in the long-time evolution of the slowly varying, ``semiclassical'', random initial data characterized  by the PDF $\mathcal{P}_0(\rho)$. 
This is done by employing the soliton gas resolution conjecture combined with the stochastic version of the IST  detailed below.

Within the IST framework soliton gases are characterized by the density of states (DOS) $f(\lambda;x,t)$---the joint distribution of solitons in the gas with respect to their Lax spectrum and their spatial positions \cite{suret_soliton_2024}. For spatially uniform soliton gases $\partial_t f = \partial_x f \equiv 0$ so that $f \equiv f(\lambda)$. For the fNLSE bound state soliton gas $f(\lambda) = \tilde f(\eta) \delta(\xi)$; we shall drop the tilde in what follows. 
 
The detailed numerical verification of the soliton gas resolution conjecture  for the IT realized in the long time development of the noise-induced modulational instability of a fNLSE plane wave solution $\psi(x,0)=\sqrt{\rho_0}>0$ was undertaken in \cite{gelash2019bound}. It was shown that observed parameters of this IT, in particular the Gaussian single-point statistics for the wave amplitude $|\psi|$, are accurately described by the statistical characteristics of the bound state soliton gas with the DOS  given by the Weyl distribution $f(\eta) = \eta/(\pi \sqrt{\rho_0-\eta^2})$ ---the so-called genus zero soliton condensate \cite{el_spectral_2020} (see  Appendix \ref{sec:condensate} for the background on soliton condensates).  { Explicitly, the results of \cite{gelash2019bound} imply the following correspondence between the PDF of the wave field intensity in the  developed IT  and the DOS of the underlying soliton gas}:
\begin{equation}
\label{eq:cond1}
\begin{split}
&f(\eta) = \frac{\eta}{\pi \sqrt{\rho_0-\eta^2}},\quad \eta\in [0,\sqrt{\rho_0}) \quad  \\
 &\mapsto \; \PDF(\rho) = \frac{1}{\rho_0} e^{-\rho/\rho_0},\quad \rho \in [0,\infty), \quad \rho_0>0.
 \end{split}
\end{equation} 
 
We take advantage of the soliton condensate DOS-PDF correspondence \eqref{eq:cond1} to realize the  stochastic analogue of the IST  procedure for the semiclassical evolution \eqref{fNLSE}, \eqref{madelung}, \eqref{slow_var}, which we schematically represent as \cite{congy_statistics_2024}
\begin{equation} \label{s_ist}
\mathcal{P}_0(\rho) \xmapsto{\text{direct SST}} f(\eta) \xmapsto{\text{inverse SST}} \mathcal{P}_\infty(\rho),
\end{equation}
where $\PDF_\infty$ is the PDF of the fully developed, spatially uniform IT at $t \to \infty$ and ``SST'' stands for ``Stochastic Scattering Transform''.
The key ingredient in the stochastic IST scheme \eqref{s_ist} is the time invariance of the DOS due to isospectrality of the fNLS evolution whereas  the time evolution step of the traditional IST is replaced  by the soliton phase randomization as $t \to \infty$. The latter can be described as an asymptotic equilibration towards the generalized Gibbs ensemble within the generalized hydrodynamics (GHD) framework \cite{doyon_lecture_2020, doyon_generalized_2025}. We also mention the GHD result  in \cite{Koch_2022} which corroborates the soliton resolution conjecture for fNLSE.

The direct SST for the semiclassical ``solitonic IT'' developing from the slowly varying random data \eqref{madelung}, \eqref{slow_var} with $\min \rho_0(x)=0$  and characterized by the PDF $\mathcal{P}_0(\rho)$ was shown in \cite{congy_statistics_2024} to  reduce to the integral Abel transform
\begin{equation}\label{d_sst}
  f(\eta) = \int_{\eta^2}^\infty
  \frac{\eta}{\pi \sqrt{\rho - \eta^2}} \PDF_0(\rho) d\rho,\quad \eta \in  [0,\infty).
\end{equation}
The generalization of this result to   the {composite soliton gas involving finite condensate background} will be presented later. 

We now proceed with the construction of the inverse SST. {The goal is to express $\PDF_\infty (\rho)$ in terms of the spectral DOS $f(\eta)$.  Motivated by the direct SST \eqref{d_sst} we introduce an ansatz}
\begin{equation}
\label{eq:transfo}
\PDF_\infty(\rho) = \int_\Lambda f(\eta) h(\eta,\rho) d\eta,
\end{equation}
where  $\Lambda$ is the spectral support of the DOS $f(\eta)$ and the kernel $h(\eta,\rho)$ is to be determined.

We take advantage of 
the benchmark DOS-PDF pair \eqref{eq:cond1} which  yields, upon substitution in \eqref{eq:transfo}, the integral equation for $h(\eta, \rho)$:
\begin{equation}
\label{eq:transfo_ex}
\int_0^{\sqrt{\rho_0}} \frac{\eta}{\pi \sqrt{\rho_0-\eta^2}} h(\eta,\rho) d\eta = \frac{1}{\rho_0} e^{-\rho/\rho_0} ,
\end{equation}
which is readily solved using the Laplace transform  to give (see Appendix \ref{sec:detail} for details):
\begin{equation}\label{eq:transfo_ex2}
\begin{split}
&    h(\eta,\rho) = \frac{1}{\eta^3} {\cal H}\left( \frac{\rho}{\eta^2}\right),\\
&    {\cal H}(t) = 2\sqrt{\frac{\pi}{t}}e^{-t/2} \,{\rm W}_{1,0}(t),
    \end{split}
\end{equation}
where ${\rm W}_{n,m}$ is the Whittaker function \cite{abramowitz_handbook_1972}. 

Substituting \eqref{eq:transfo_ex2} in \eqref{eq:transfo} we obtain the general formula for the inverse SST yielding the PDF of the wave intensity in  the bound state soliton gas with a given DOS $f(\eta)$:
\begin{equation}
\label{eq:transfo2}
 \PDF_\infty(\rho) =\int_\Lambda \frac{f(\eta)}{\eta^3}  {\cal H}\left( \frac{\rho}{\eta^2}\right) d\eta . \\
\end{equation}
The DOS in \eqref{eq:transfo2} is subject to the PDF normalization, $\int_0^\infty \PDF_\infty(\rho) d\rho=1$. This implies that for any  soliton gas with the DOS' support $\Lambda = [0,\beta)$ (including $\beta \to \infty$), the following constraint for the DOS must hold 
\begin{equation} \label{cond_0}
\int_0^\beta \frac{2f(\mu)}{\mu} d\mu= 1,
\end{equation}
obtained using $\int_0^\infty {\cal H}(t) dt=2$. {Note that the same condition \eqref{cond_0} follows from inverting the Abel transform \eqref{d_sst} and using the normalization $\int_0^\infty\PDF_0(\rho)d \rho=1$.}

{The inverse SST relation \eqref{eq:transfo2} provides the basis for the main analytical results of the paper, which will be verified by direct numerical simulations across several representative examples of fNLSE IT. We stress that this relation is valid for any bound state soliton gas satisfying condition \eqref{cond_0}, see the End Matter for an instructive  example.
Although the normalization \eqref{cond_0} might appear quite restrictive,  we show below that it is satisfied for a broad class of physically relevant bound state soliton gases. In what follows we explore the physically important consequences of eq.~\eqref{eq:transfo2} and consider its application to the statistical characterization of semiclassical IT. }

Multiplying \eqref{eq:transfo2} by $\rho^\alpha$, $\alpha >0$, and integrating $\int_0^{\infty} \dots d\rho$ we obtain the relation between the moments of the PDF and the moments of the DOS in the semiclassical IT:
\begin{equation}\label{mom}
\begin{split}
\langle \rho^\alpha\rangle &=\int_0^\infty \rho^\alpha\, \PDF_\infty(\rho) d\rho \\ 
&= \frac{2 \sqrt \pi \,\Gamma^2(\alpha+1)}{\Gamma(\alpha+1/2)} \int_\Lambda \eta^{2\alpha-1} f(\eta) d\eta,
\end{split}
\end{equation}
where $\Gamma(z)$ is Euler's Gamma function. Note that $\alpha$ should not necessarily be an integer. {Thus, the IT framework enables the determination of  all statistical moments {\it without imposing any closure assumptions} of the kind required in classical turbulence theories \cite{pope_turbulent_2000, nazarenko_wave_2011}.} From \eqref{mom} we notably recover  the formula for the  fourth normalized moment of $|\psi|$ (associated with the kurtosis of $\mathcal{P}(\rho)$) of the bound state soliton gas derived in \cite{congy_statistics_2024} (this result was also obtained in \cite{Koch_2022} within the GHD framework via the semiclassical limit of the  quantum Bose gas):
\begin{equation}\label{kurt}
  \kappa_4=  \frac{\int_0^\infty \rho^2\, \PDF_\infty(\rho) d\rho}{\left(\int_0^\infty \rho \PDF_\infty(\rho) d\rho\right)^2}  
  = \frac{2}{3} \frac{\int_\Lambda \eta^{3} f(\eta) d\eta} {\left(\int_\Lambda \eta f(\eta) d\eta \right)^2}.
\end{equation}
Similarly, we obtain for the third normalized moment of $|\psi|$, associated with the skewness of $\mathcal{P}(\rho)$:
\begin{equation}\label{skew}
  \kappa_3=  \frac{\int_0^\infty \rho^{3/2}\, \PDF_\infty(\rho) d\rho}{\left(\int_0^\infty \rho \PDF_\infty(\rho) d\rho\right)^{3/2}}  
  =\frac{9 \pi^{3/2}}{64}\frac{\int_\Lambda \eta^{2} f(\eta) d\eta} {\left(\int_\Lambda \eta f(\eta) d\eta \right)^{3/2}}.
\end{equation}
{Both kurtosis and skewness are important predictors for the presence of rogue waves in nonlinear random wave fields \cite{onorato_freak_2001, kharif2008Rogue}. We note that the skewness formula \eqref{skew} has not been previously available; here it is obtained simply as a consequence of the general result \eqref{mom}.}

We now apply the inverse SST formula \eqref{eq:transfo2}  to the {particular} case of IT emerging in the long-time evolution of narrow-band partially coherent waves  with exponential PDF $\PDF_0(\rho) = \rho_0^{-1}e^{-\rho/\rho_0}$ at $t=0$, numerically studied in \cite{agafontsev_extreme_2021} and realized in fiber optics experiments \cite{walczak_optical_2015, suret_single-shot_2016}. The direct SST \eqref{d_sst}  yields, for $\rho_0=1$, the Rayleigh distribution for the DOS 
\begin{equation}\label{rayleigh}
    f(\eta) = \frac{\eta  \exp \left(-\eta ^2\right)}{\sqrt{\pi }},\quad \eta \in [0,\infty),
\end{equation}
which satisfies the normalization \eqref{cond_0}. 
Substituting \eqref{rayleigh} into \eqref{eq:transfo2} we obtain
\begin{equation}\label{eq:pdf_pc}
    \PDF_\infty(\rho) =  2\Kt_0 (2\sqrt{\rho}),
\end{equation}
where $\Kt_n$ is the modified Bessel function of the second kind \cite{abramowitz_handbook_1972}, in agreement with the heuristic result of \cite{agafontsev_extreme_2021}  verified by direct  numerical simulations of fNLSE.

We next apply the stochastic IST setting \eqref{s_ist} to determine the PDF $\PDF_\infty (\rho)$ for a more general class of IT {identified  in \cite{biondini_breather_2025} by considering the semiclassical fNLSE evolution of the elliptic $\mathrm{dn}$ potential modified by weak random perturbation}.
{Motivated by this example} we consider fNLSE with the initial condition
\begin{equation}\label{init_cond}
    \psi(x, 0) = \sqrt{\rho_0(x)} + \varepsilon(x),\quad \rho_0(x)>0,
\end{equation}
where $\rho_0(x) \in [\rho_-,\rho_+]$ is a $L$-periodic  single-lobe function with $L \gg 1$, satisfying the  slow variation condition \eqref{slow_var},
and $\varepsilon(x) \in \mathbb{C}$ is a small random noise with zero mean, $|\varepsilon(x)| \ll 1$, $\langle \varepsilon(x) \rangle =0$. {As shown in \cite{biondini_breather_2025} the sufficiently small complex valued random perturbation does not affect the DOS of the potential $\sqrt{\rho_0(x)}$ at leading order but introduces small non-zero velocities to the constituent bound state solitons, facilitating  the soliton gas resolution as $t \to \infty$.}

Neglecting the effect of the small perturbation $\varepsilon(x)$ on the spectrum of $\sqrt{\rho_0(x)}$ the time-invariant DOS of \eqref{init_cond} can be evaluated as follows  \cite{biondini_semiclassical_2020, tovbis_recent_2022, biondini_breather_2025}:
\begin{equation} \label{pdf_period}
\begin{split}
   & f(\eta)  =\\
   & \begin{cases}
    \displaystyle
            \frac{1}{L} \int_0^L \frac{\eta}{\pi \sqrt{\rho_0(x)-\eta^2}} dx, 
            & 0 \leq \eta < \sqrt{\rho_-},\\[5mm]
    \displaystyle        
            \frac{1}{L} \int_{R_0(\eta)} \frac{\eta}{\pi \sqrt{\rho_0(x)-\eta^2}} dx, & \sqrt{\rho_-} \leq \eta < \sqrt{\rho_+},    
            \end{cases}
              \end{split}
\end{equation} 
where $R_0 (\eta)\subset [0,L]$ is the region of space such that $\rho_0(x)>\eta^2$.

Upon replacing $L$ with  $\ell \gg L$,  introducing the change of variable $r=\rho_0(x)$,   and passing to the limit $\ell \to \infty$ equation \eqref{pdf_period} becomes 
\begin{equation}
\label{eq:mod}
\begin{split}
   & f(\eta) = \\
   & \begin{cases}
    \displaystyle
            \int_{\rho_-}^{\rho_+} \frac{\eta}{\pi \sqrt{r-\eta^2}} \,\PDF_0(r) dr, 
            & 0 \leq \eta < \sqrt{\rho_-},\\[5mm]
    \displaystyle        
            \int_{\eta^2}^{\rho_+} \frac{\eta}{\pi \sqrt{r-\eta^2}} \,\PDF_0(r) dr, & \sqrt{\rho_-} \leq \eta < \sqrt{\rho_+},
              \end{cases}
              \end{split}
\end{equation}
where 
\begin{equation}\label{pdf0}
 \PDF_0(\rho) = \lim \limits_{\ell \to \infty}\frac{1}{\ell} \sum_i \left| \frac{d x_i}{d\rho} \right|, 
\end{equation}
with $x_i \in [0, \ell]$ satisfying $\rho_0(x_i) = \rho$.
The function $\PDF_0(\rho)$ defined by \eqref{pdf0} 
is the PDF of the initial periodic function $\rho_0(x)$ interpreted  as an ergodic random process by adding to the argument a random variable $\omega$ uniformly distributed over the circle $[0,L)$ \cite{pastur_spectra_1992}.
We now observe  that the  probabilistic representation  \eqref{eq:mod} of the DOS,  unlike the direct semiclassical formula \eqref{pdf_period}, is not restricted to periodic data and  {represents the generalization of the direct SST \eqref{d_sst}  to semiclassical random potentials with non-zero background. Indeed, for $\rho_-=0$, $\rho_+ \to \infty$  formula \eqref{eq:mod} reduces to the Abel transform \eqref{d_sst}.}

\begin{figure*}
    \centering
    \includegraphics[width=0.9\linewidth]{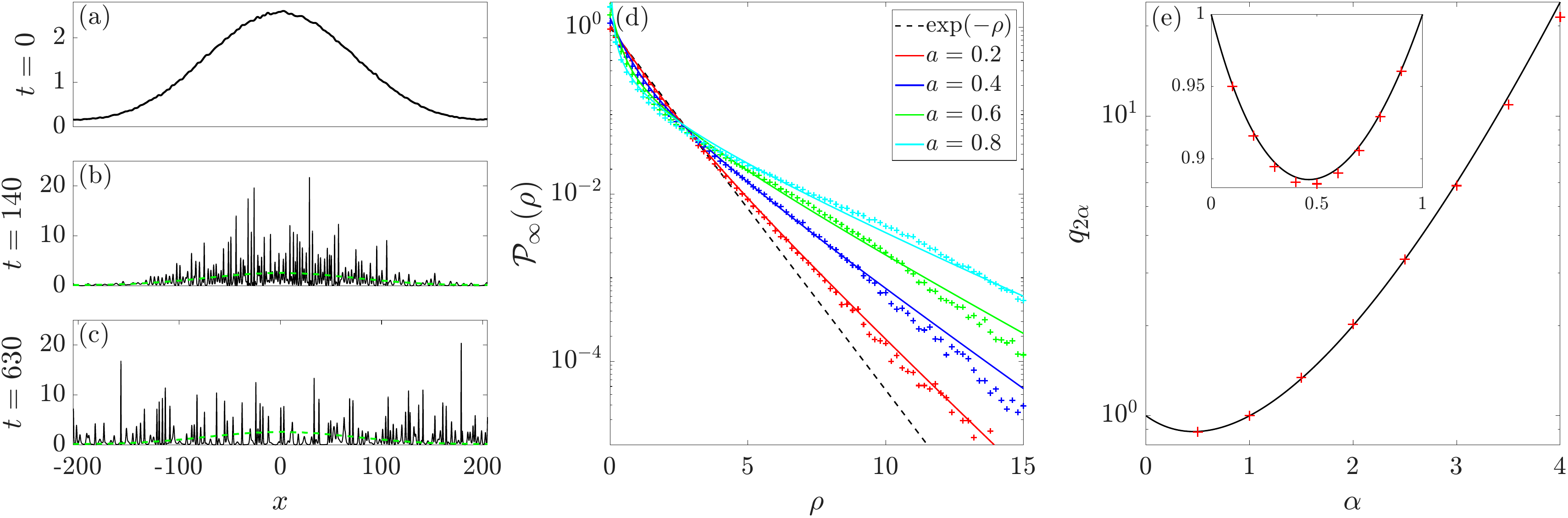}
    \caption{Integrable turbulence emerging from the the periodic initial state \eqref{eq:initcos} modified by a small random noise. (a-c) numerical solution $|\psi(x,t)|^2$ of \eqref{fNLSE}, \eqref{init_cond}, \eqref{eq:initcos} (solid black line)  with $a=0.6$ and $L= 4.096 \times 10^2$. For reference the dashed green line represents the initial condition in (b) and (c);
    (d) comparison between the numerically evaluated PDF for several values of $a$ (color markers) and the corresponding  PDF evaluated using Eqs. \eqref{eq:pdf_inf1}, \eqref{eq:cos} (solid lines). The dashed black line represents the exponential PDF $e^{-\rho}$ (Gaussian single-point statistics for the fNLSE field amplitude);
    (e) comparison between the numerically evaluated moment ratio $q_{2\alpha}$\eqref{eq:gamma} {for $a=0.6$} (red markers) and $\Gamma(\alpha+1)$ (solid line). The inset shows a magnified view of the region $\alpha \in [0,1]$. The details of the numerical computations are given in Appendix \ref{sec:numerics}.
    }
    \label{fig:cos}
\end{figure*}

The DOS \eqref{eq:mod} corresponds to a semiclassical {\it composite soliton gas}, i.e. a ``regular'' semiclassical soliton gas on the background of a {\it soliton condensate} \cite{biondini_breather_2025}. { Such gas  can also be interpreted as breather gas \cite{el_spectral_2020, roberti_numerical_2021} following the conventional notion of breathers as solitons on finite background \cite{akhmediev1986modulation, Kibler2010Peregrine, Chabchoub2011Rogue}. The concept of composite soliton gas  was instrumental   for the description of the wave-mean field interaction in the KdV IT \cite{congy_exactly_2026}.}

Spectrally, a composite soliton gas is defined as follows (see Appendix \ref{sec:condensate} for the additional background material).  Let $\Lambda = [0, \sqrt{\rho_+})$, where the background soliton condensate component of the composite gas is supported on $\eta \in [0, \sqrt{\rho_-})$, where $\rho_-<\rho_+$. Then the DOS of a composite soliton gas  satisfies the following  condition:
\begin{equation}
\label{eq:ndr_cond}
    \int_0^{\sqrt{\rho_+}} G(\eta,\mu) f(\mu) d\mu  = 1,\quad  \eta \in [0,\sqrt{\rho_-}),
\end{equation}
where $G(\eta, \mu) = \eta^{-1} \ln |(\mu+ \eta)/(\mu - \eta)|$ is the two-soliton position shift kernel \cite{Zakharov:1972Exact}. 

Substitution of \eqref{eq:mod} into \eqref{eq:ndr_cond} confirms
that
condition \eqref{eq:ndr_cond} is indeed satisfied (see Appendix \ref{sec:condensate} for the details of the computation). In fact, Eq.~\eqref{eq:ndr_cond} represents a generalization of the normalization  \eqref{cond_0}  obtained for ``regular'' (non-composite) gases.
Indeed, in the particular case $\rho_-=0$,
 we evaluate
$\lim_{\eta \to 0} \int_\Lambda G(\eta,\mu) f(\mu) d\mu = \int_\Lambda \lim_{\eta \to 0}  G(\eta,\mu)  f(\mu) d\mu = \int_\Lambda \frac{2f(\mu)}{\mu} d\mu$
so that
condition \eqref{eq:ndr_cond} assumes the form \eqref{cond_0} with $\beta = \sqrt{\rho_+}$.

Substituting \eqref{eq:mod} in \eqref{eq:transfo2}   we obtain the direct relation between $\mathcal{P}_0$ and $\mathcal{P}_\infty$  in  semiclassical IT (see Appendix \ref{sec:detail} for details of the derivation):
\begin{equation}
\label{eq:pdf_inf1}
\PDF_\infty(\rho)  = \int_{\rho_-}^{\rho_+} \PDF_0(r) \, \frac{e^{-\rho/r}}{r} dr .
\end{equation}
One can readily verify the PDF normalization $\int_0^\infty \mathcal{P}_\infty(\rho)d \rho =1$. {A particular case of the general relation \eqref{eq:pdf_inf1} appears in \cite{agafontsev_extreme_2021} for $\rho_-=0$, $\rho_+ \to \infty$, $\mathcal{P}_0(\rho)= e^{-\rho}$, corresponding to the IT developing from Gaussian narrow-band partially coherent waves. }

Next,  using \eqref{eq:pdf_inf1} we obtain a remarkably simple universal relation between the moments of the PDF at $t=0$ and $t\to \infty$:
\begin{equation}
\label{eq:gamma}
    q_{2\alpha}= \dfrac{\int_0^\infty \rho^\alpha\, \PDF_\infty(\rho) d\rho}{\int_{\rho_-}^{\rho_+} \rho^\alpha \,\PDF_0(\rho) d \rho} = \Gamma(\alpha+1),
\end{equation}
{which for $\alpha=2$ yields the  kurtosis doubling result  obtained for the regular semiclassical soliton  gases in \cite{congy_statistics_2024} and for the ``elliptic'' composite gas in  \cite{biondini_breather_2025}. It has been subsequently verified in a fiber optics experiment \cite{copie_experimental_2025}.}

\medskip
Motivated by the theoretical and experimental results of \cite{biondini_breather_2025, copie_experimental_2025}   we consider the problem of the {soliton gas resolution} for the fNLSE  with initial data \eqref{init_cond}, where
\begin{equation}
\label{eq:initcos}
  \sqrt{\rho_0(x)} = 1+a\cos \left( \frac{2\pi}{L} \,x\right), \  \ x \in [0,L],
\end{equation}
with $L \gg 1$ and $0<a<1$.
A typical realization of the  IT field developing from the initial data \eqref{init_cond}, \eqref{eq:initcos} is shown in Fig.~\ref{fig:cos}(a)-(c). 

The PDF of the initial distribution \eqref{eq:initcos} is evaluated by \eqref{pdf0} to give:
\begin{equation}
\label{eq:cos}
\begin{split}
   & \PDF_0(\rho) = \frac{1}{2\pi\sqrt{\rho(1+a-\sqrt \rho)(\sqrt \rho-1+a)}},\\ & \rho \in \left((1-a)^2,(1+a)^2 \right).
    \end{split}
\end{equation}

Substituting 
\eqref{eq:cos} in \eqref{eq:pdf_inf1} we obtain the PDF of the wave field intensity in the fully developed breather IT at $t \to \infty$. {An explicit analytical expression for $\PDF_\infty(\rho)$ is not available but the integral in \eqref{eq:pdf_inf1} with $\PDF_0(\rho)$ \eqref{eq:cos} is readily amenable to the numerical evaluation.} The plots of $\PDF_\infty(\rho)$ for different values of $a$ are presented in Fig.~\ref{fig:cos}(b) and compared with the PDF extracted from fNLSE numerical simulations. One can see a very good agreement between the analytical and numerical results, {which is expected to further improve with the increase of the evolution time and the spatial period $L$ leading to further spatial uniformization of the gas and better accuracy of the semiclassical approximation respectively, see Appendix \ref{sec:numerics}.}
In Fig.~\ref{fig:cos}(e) we present comparison of the analytically predicted moment ratio $q_{2\alpha}$ given by Eq.~\eqref{eq:gamma} with  the moment ratio extracted from direct  numerical simulations of fNLSE with initial condition \eqref{eq:initcos} modified by small random noise (cf. \eqref{init_cond}). Again, an excellent agreement is observed. The details of the numerical  simulations are presented in Appendix \ref{sec:numerics}.

Evolution \eqref{fNLSE}, \eqref{init_cond}, \eqref{eq:initcos} was realized in the fiber optics experiment \cite{copie_experimental_2025}, where the  doubling of the kurtosis was observed in agreement with \eqref{eq:gamma}.

{In conclusion, we formulated the soliton gas resolution conjecture and  determined the probability density function (PDF) for the wave field intensity  in   fNLSE integrable turbulence approximated by a bound state soliton gas.
For a given spectral density of states (DOS) the PDF is determined by the inverse stochastic spectral transform \eqref{eq:transfo2}, which is the key analytical result of our work enabling concrete applications.
We employed this general analytical result to obtain the PDF for several example cases of  semiclassical integrable turbulence emerging in the long-time  evolution of slowly varying  random initial data with excellent agreement observed between the theory and direct  numerical simulations of fNLSE integrable turbulence. Notably, our analytic results are exact and don't involve any ad hoc closure assumptions, in contrast to classical turbulence theories.}

{This work represents an important step toward complete analytical characterization of nonlinear random wave fields in physical systems modelled by integrable dispersive wave equations. The next step will be the determination of the correlation function and Fourier power spectrum of integrable turbulence, the key observables in physical experiments \cite{suret_soliton_2024}.}  Future directions also include the generalization of the developed theory to non-bound state soliton gases 
and  other integrable systems. Further, in weakly non-uniform soliton gases  the DOS and hence, the PDF, depend on space and time. In such setting, the stochastic IST \eqref{s_ist} provides an IT description at the local-equilibrium, “mesoscopic” scale, and it should therefore be complemented by a spectral kinetic equation governing the evolution of the DOS on the macroscopic (Euler) scale \cite{el_kinetic_2005, el_spectral_2020, el_soliton_2021}.

\begin{acknowledgments}
This work was supported by the EPSRC grant No. EP/C002401/1.  We thank G. Biondini, M. Hoefer,  S. Randoux, G. Roberti, P. Suret, and A. Tovbis for many fruitful discussions.
\end{acknowledgments}

\appendix

\onecolumngrid

\section{Soliton condensate and composite soliton gas}
\label{sec:condensate}

The DOS of the bound state fNLSE soliton gas satisfies the nonlinear dispersion relation (NDR) \cite{el_spectral_2020}
\begin{equation}\label{ndr1}
\int_{\Gamma} G(\eta,\mu) f(\mu) \, d\mu +  \frac{\sigma(\eta)}{\eta} f(\eta)
= 1,
\end{equation}
where $G(\eta, \mu) = \eta^{-1} \ln |(\mu+ \eta)/(\mu - \eta)|$ is the two-soliton position shift kernel \cite{Zakharov:1972Exact}, $\Gamma$ is the DOS spectral support, and  $\sigma(\eta)>0$ is the so-called spectral scaling function appearing in the  soliton gas construction via the thermodynamic limit of finite-gap potentials \cite{el_spectral_2020}. 

Soliton condensates represent critically dense soliton gases in which collective soliton interactions supersede the dynamics of individual solitons \cite{el_spectral_2020, congy_dispersive_2023}. Spectrally  soliton condensates are defined  by the condition $\sigma(\eta)\to 0$ in the NDR \eqref{ndr1}, which yields the integral equation for the condensate DOS 
\begin{equation}
\label{eq:ndr_cond0}
    \int_\Gamma G(\eta, \mu) f(\mu) d\mu  = 1.
\end{equation}
Let $\Gamma =[0,\sqrt{\rho_-})$. Then equation \eqref{eq:ndr_cond0} is solved by \cite{el_spectral_2020} (cf. Eq.~\eqref{eq:cond1})
\begin{equation}\label{weyl}
f(\eta) = \frac{\eta}{\pi \sqrt{\rho_--\eta^2}} \equiv f_c(\eta).
\end{equation}

The notion of a composite soliton gas was introduced in \cite{biondini_breather_2025} as a soliton gas consisting of  condensate and non-condensate components. Importantly, the condensate component includes the induced mean field from the non-condensate gas. As a result, the DOS of a composite soliton gas, supported on $\Lambda=[0, \sqrt{\rho_+}), \ \rho_+>\rho_-$, admits the following partitioning  \cite{congy_exactly_2026}:
\begin{equation}
\label{eq:composite}
f(\eta) = \underbrace{f_c(\eta) - f_{\text{IMF}}(\eta)}_{\mathrm{condensate}} \hspace{2mm}+ \hspace{-5mm}\underbrace{\ft(\eta)}_{\mathrm{non-condensate}},
\end{equation}
where the \textit{core} condensate DOS $f_c(\eta)$ and the {\it induced mean field} condensate DOS $f_\text{IMF}(\eta)$ are both supported on $\Gamma =[0, \sqrt{\rho_-}) \subset \Lambda$,  while the non-condensate DOS $\ft(\eta)$ is supported on ${[\sqrt{\rho_-}, \sqrt{\rho_+})}$.  As shown in \cite{congy_exactly_2026} the DOS decomposition \eqref{eq:composite} enables the effective separation of the mean field (condensate) from the stochastic, fluctuating solitons (non-condensate).
 Then the DOS of a composite soliton gas  satisfies the following  condition (Eq.~\eqref{eq:ndr_cond})
\begin{equation}
    \int_\Lambda G(\eta,\mu) f(\mu) d\mu  = 1,\quad  \eta \in \Gamma\, .
\end{equation}

\section{Details of some derivations}
\label{sec:detail}

{\bf Solution of the integral equation~\eqref{eq:transfo_ex}}

\medskip

Introducing the change of variable $t=\eta^2$
and applying the Laplace transform $\mathcal{L}[\cdot]$ with respect to $\rho_0$, we obtain

\begin{equation}
\frac{1}{2\sqrt {\pi s}} H(s,\rho) = 2{\rm K}_0(2\sqrt{s\rho}), \quad H(s,\rho)={\cal L}[h(\sqrt t,\rho)] = \int_0^\infty h(\sqrt t,\rho) e^{-st} dt,
\end{equation}
where ${\rm K}_n$ is the modified Bessel function of the second kind \cite{abramowitz_handbook_1972}.
Computing the inverse Laplace transform of $H(s,\rho) = 4\sqrt {\pi s} \, {\rm K}_0(2\sqrt{s\rho})$ yields Eq.~\eqref{eq:transfo_ex2}.

\

{\bf Verification of the composite soliton gas condition \eqref{eq:ndr_cond}}

\medskip
We have, using the DOS of semi-classical soliton gas \eqref{eq:mod}
\begin{equation} \label{condition}
\begin{split}
   \int_0^{\sqrt{\rho_+}} G(\eta,\mu) f(\mu) d\mu &= \int_0^{\sqrt{\rho_-}} G(\eta,\mu) \left( \int_{\rho_-}^{\rho_+} \frac{\mu}{\pi \sqrt{r - \mu^2}} \PDF_0(r) d r \right) d\mu\\
    &\quad +\int_{\sqrt{\rho_-}}^{\sqrt{\rho_+}} G(\eta,\mu) \left( \int_{\mu^2}^{\rho_+} \frac{\mu}{\pi \sqrt{r - \mu^2}} \PDF_0(r) d r \right) d\mu\\
  &  = \int_{\rho_-}^{\rho_+} \PDF_0(r) \left( \int_0^{\sqrt{r}} \frac{\mu}{\pi \sqrt{r - \mu^2}} G(\eta,\mu) d\mu\right) dr 
     = \int_{\rho_-}^{\rho_+} \PDF_0(r) dr=1,
\end{split}
\end{equation}
where we have used  the identity
\begin{equation}
\int_0^{\sqrt{r}} \frac{\mu}{\pi \sqrt{r - \mu^2}} G(\eta,\mu) d\mu= 
        1,\quad \forall \eta \leq \sqrt r.
\end{equation}
Hence the DOS \eqref{eq:mod} indeed defines a composite soliton gas.
We can thus partition this DOS  as in \eqref{eq:composite}
\begin{equation}
\label{eq:composite2}
    f(\eta) = \begin{cases}
    \displaystyle
            \int_{\rho_-}^{\rho_+} \frac{\eta}{\pi \sqrt{r-\eta^2}} \,\PDF_0(r) dr = f_c(\eta) - f_{\text{IMF}}(\eta), 
            & 0 \leq \eta < \sqrt{\rho_-},\\[5mm]
    \displaystyle        
            \int_{\eta^2}^{\rho_+} \frac{\eta}{\pi \sqrt{r-\eta^2}} \,\PDF_0(r) dr \equiv \ft(\eta), & \sqrt{\rho_-} \leq \eta < \sqrt{\rho_+},
              \end{cases}
\end{equation}
where $f_c(\eta)$ is the condensate DOS defined in \eqref{weyl} and the induced mean field condensate DOS is given by \cite{congy_exactly_2026}:
\begin{equation}
\label{eq:imf}
    f_{\rm IMF}(\eta) = \frac{\eta}{\pi \sqrt{\rho_- -\eta^2}} \int_{\sqrt{\rho_-}}^{\sqrt{\rho_+}} \frac{2 \sqrt{\mu^2-\rho_-}}{\mu^2 -\eta^2} \ft(\mu) d\mu,\quad \eta \in [0,\sqrt{\rho_-}).
\end{equation}
The equality for $\eta < \sqrt{\rho_-}$ in \eqref{eq:composite2} is obtained by exchanging the order of integration, similarly to the computation in \eqref{condition}.

\

{\bf Derivation of Eq.~\eqref{eq:pdf_inf1}}

\medskip
Substituting Eq.~\eqref{eq:mod} in \eqref{eq:transfo2}, we obtain:
\begin{equation}
\label{eq:pdf}
\begin{split}
    \PDF_\infty(\rho) & = \int_0^{\sqrt{\rho_-}} \frac{1}{\eta^3}  {\cal H}\left( \frac{\rho}{\eta^2}\right) \left( \int_{\rho_-}^{\rho_+} \frac{\eta}{\pi \sqrt{r - \eta^2}} \PDF_0(r) d r \right) d\eta
    +\int_{\sqrt{\rho_-}}^{\sqrt{\rho_+}} \frac{1}{\eta^3}  {\cal H}\left( \frac{\rho}{\eta^2}\right) \left( \int_{\eta^2}^{\rho_+} \frac{\eta}{\pi \sqrt{r - \eta^2}} \PDF_0(r) d r \right) d\eta \\
    & = \int_{\rho_-}^{\rho_+} \PDF_0(r) \left( \int_0^{\sqrt{r}} \frac{\eta}{\pi \sqrt{r - \eta^2}} \frac{1}{\eta^3}  {\cal H}\left( \frac{\rho}{\eta^2}\right) d\eta\right) dr 
     = \int_{\rho_-}^{\rho_+} \PDF_0(r) \, \frac{e^{-\rho/r}}{r} dr,
\end{split}
\end{equation}
where the last equality is obtained via the integral representation of ${\cal H}$ by Eqs. \eqref{eq:transfo_ex}, \eqref{eq:transfo_ex2}.

\section{End Matter}

\begin{figure}
    \centering
    \includegraphics[width=0.3\linewidth]{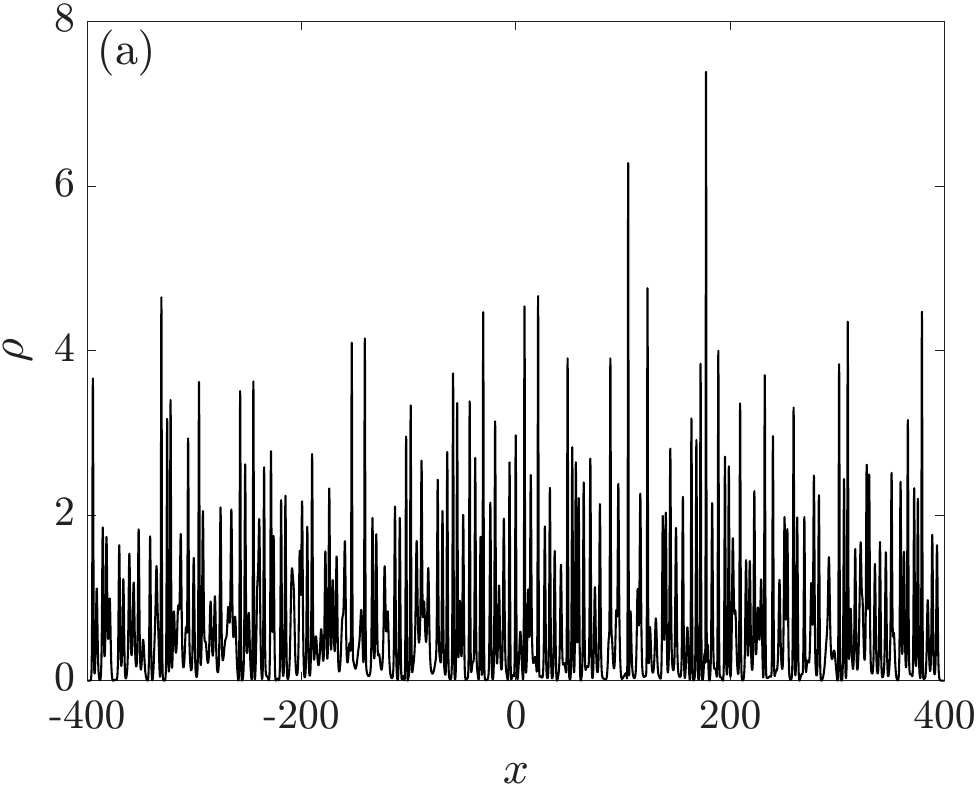}
\includegraphics[width=0.3\linewidth]{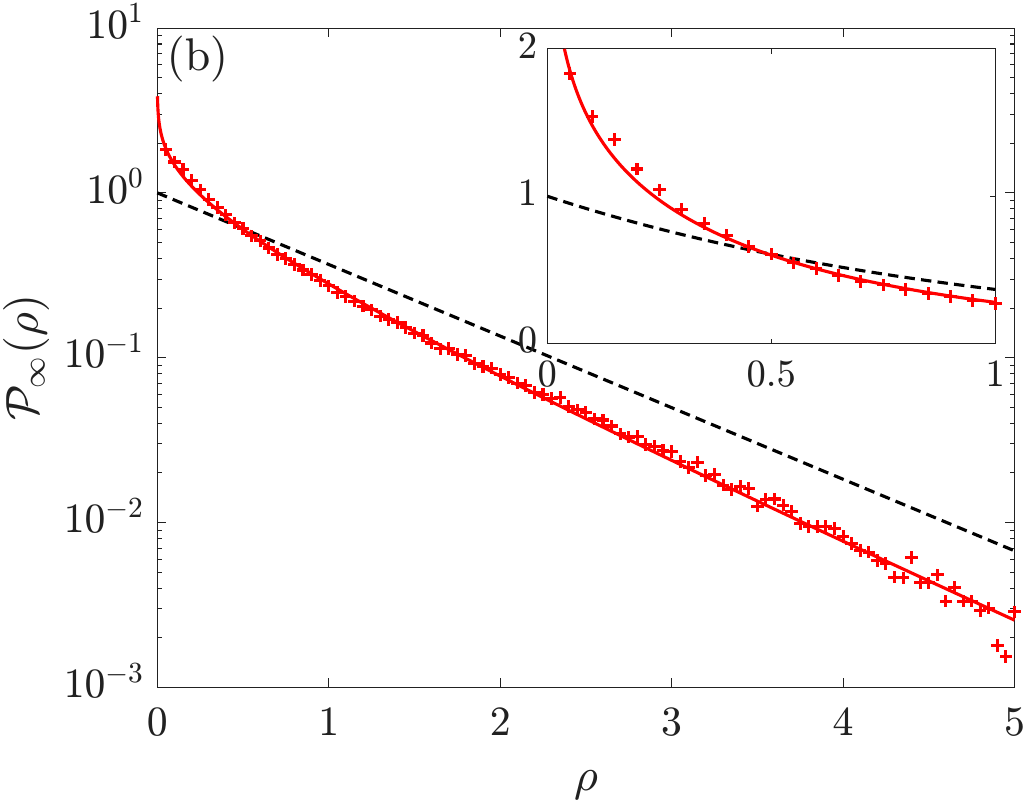}
    \caption{{Soliton gas with the linear DOS \eqref{linear}. (a) Typical realization modeled by a random $n$-soliton solution with $n=200$.  (b) PDF of $\rho=|\psi|^2$. The inset plot displays a zoom-in  of the region $\rho \in [0,1]$. The solid red line represents the analytical expression \eqref{linear_pdf} and the red markers the PDF computed numerically over $50$ realizations. For reference the dashed black line represents the PDF $e^{-\rho}$ corresponding to the Gaussian statistics of the wave amplitude.}
    }
    \label{fig:linear}
\end{figure}
{
In order to illustrate the applicability of the inverse SST formula \eqref{eq:transfo2} to a  bound state soliton gas, which does not necessarily emerge from a slowly varying initial condition, we now consider the  soliton gas with the linear DOS:
\begin{equation}\label{linear}
    f(\eta) = \frac{\eta}{2}, \quad \eta \in \Lambda=[0,1),
\end{equation}
which trivially fulfills the normalization condition \eqref{cond_0}. A typical  realization of such soliton gas is shown in Fig.~\ref{fig:linear}(a). Substituting the DOS \eqref{linear} in the inverse SST \eqref{eq:transfo2}, we obtain the PDF of the wave field intensity:
\begin{equation}\label{linear_pdf}
    \PDF_\infty(\rho) = \frac{e^{-\rho/2}}{2} \Kt_0 \left(\frac{\rho}{2} \right).
\end{equation}
Fig. \ref{fig:linear}(b) displays excellent agreement between the analytical expression \eqref{linear_pdf} and the PDF obtained numerically with $50$ realizations.
}

\section{Numerical methods}
\label{sec:numerics}

\subsection{Semiclassical evolution}

The initial condition implemented numerically is (Eqs. \eqref{init_cond}, \eqref{eq:initcos}) 
\begin{equation}
\label{eq:init}
    \psi(x,0) = 1+a\cos\left( \frac{2\pi}{L} x\right) + \varepsilon(x),\quad x \in \left[-\frac{L}{2}, \frac{L}{2} \right],\quad L =  4.096 \times 10^2,
\end{equation}
where the noise $\varepsilon$ is generated by a sum of incoherent, discrete Fourier components:
\begin{equation}
\label{eq:noise}
  \varepsilon(x) = \varepsilon_0 \sum_{j=1}^{P} \exp\left[-\frac{k_j^2}{\Delta k^2}+i (k_jx+\sigma_0^j)\right],
\end{equation}
with $k_j = \frac{2\pi}{L} j$, $\varepsilon_0=10^{-3}$, $\Delta k=1.5$, $P=100$ and $\sigma_0^j$ uniformly distributed between $0$ and $2\pi$. 
The fNLSE with periodic boundary conditions: $\psi(-L/2,t) = \psi(L/2,t)$ is solved via a pseudo-spectral, fourth order Runge-Kutta method \cite{trefethen_spectral_2000} with the discretization $(\Delta x,\Delta t) = (10^{-1},5\times 10^{-3} )$ and a number of points $N_x=L/\Delta x=2^{12}$. 
The statistics is then extracted from $R=10$ different numerical realizations $\psi_k(x,t)$ evolving from the random initial condition \eqref{eq:init}, \eqref{eq:noise}, in the long time regime as defined below.

\begin{figure}
    \centering
    \includegraphics[width=0.6\linewidth]{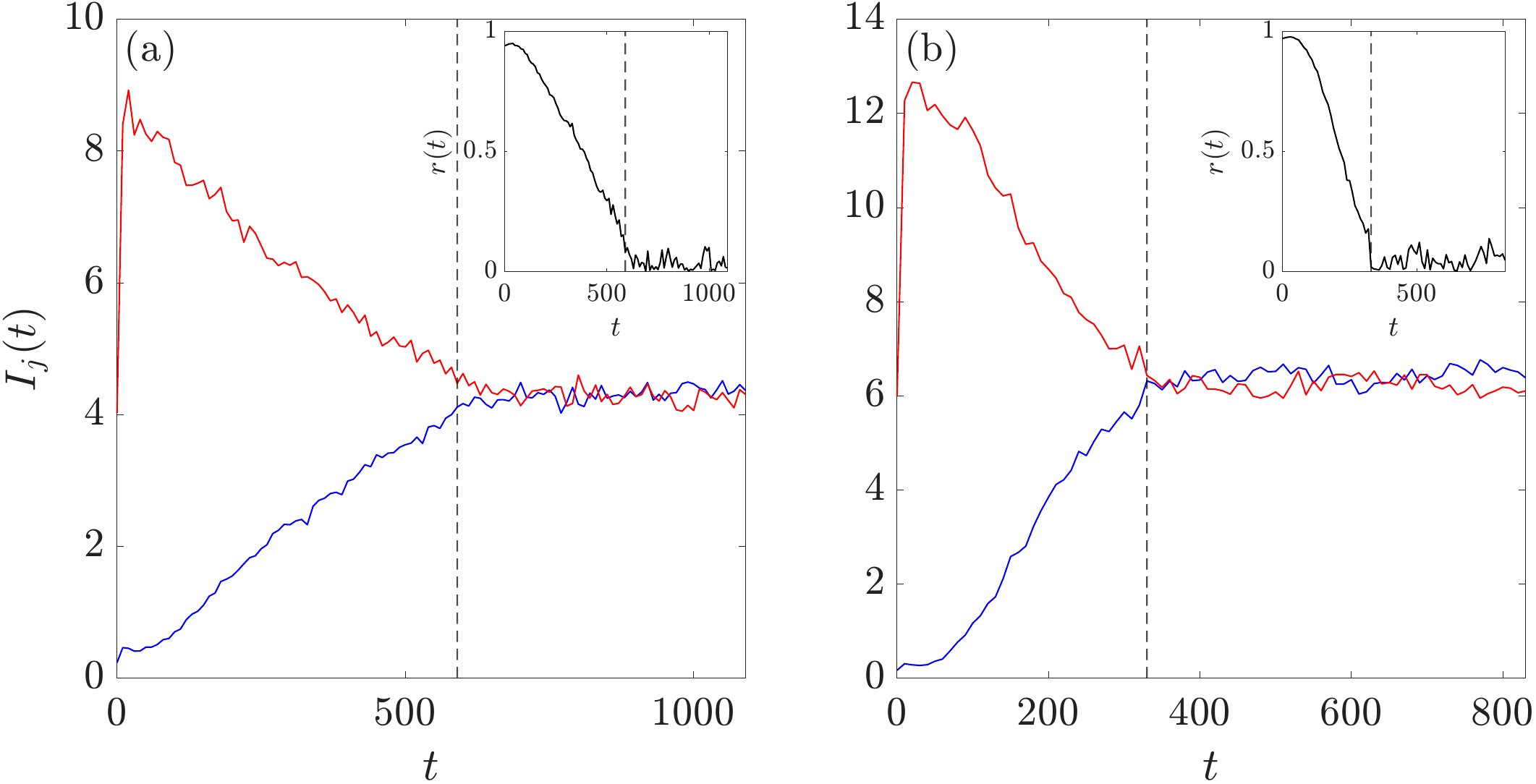}
    \caption{Evolution of spatial averages $I_1(t)$ (blue line) and $I_2(t)$ (red line) defined in \eqref{eq:local} for (a) $a=0.6$ and (b) $a=0.8$. The vertical dash line indicates the position of $\tau_a$. The inset plots show the variation of $r(t)$ defined in \eqref{eq:r}.
    }
    \label{fig:I}
\end{figure}

We evaluate the average of $|\psi|^4$ in different regions of space:
\begin{equation}
\label{eq:local}
    I_1(t) = \frac{2}{L} \int_{L/4}^{3L/4} \overline{|\psi(x,t)|^4} dx,\quad I_2(t) = \frac{2}{L} \int_{-L/4}^{L/4} \overline{|\psi(x,t)|^4} dx,
\end{equation}
where $\overline{|\psi(x,t)|^4}$ denotes the ensemble-average of the $R$ different realizations:
\begin{equation}
\label{eq:ens_aver}
    \overline{|\psi(x,t)|^{2\alpha}} = \frac{1}{R} \sum_{k=1}^R |\psi_k(x,t)|^{2\alpha}.
\end{equation}
Substituting \eqref{eq:init} in \eqref{eq:local}, \eqref{eq:ens_aver} we show that the initial profile is spatially inhomogeneous since $I_1(0)<1<I_2(0)$. Figure \ref{fig:I} shows that, for different values of $a$, $I_1(t)$ qualitatively increases and $I_2(t)$ qualitatively decreases in the early time regime. We can assume that the long-time, spatially homogeneous regime of IT is reached when the local averages are equal to each other $I_2(t) \sim I_1(t)$. We observe in Figure \ref{fig:I} that this asymptotic regime is reached at different times $t=\tau_a$ depending on the the value of the initial amplitude $a$. Numerically we define $\tau_a$ by the criterion
\begin{equation}
\label{eq:r}
    r(t)=\frac{|I_2(t)-I_1(t)|}{I_2(t)} < 1.5 \times 10^{-1},\quad \forall t \geq \tau_a.
\end{equation}

The single-point statistics (moments $q_{2\alpha}$ and PDF ${\cal P}_\infty(\rho)$) is computed from the different realizations $\psi_k(x,t)$. The ergodicity property of homogeneous soliton gases implies that ensemble-average is equal to spatial or temporal-average, and the statistics is extracted from the set of points 
\begin{equation}
    \Psi = \left\{ \psi_k(x,t) \;\big|\; (k,x,t) \in \{1,\dots,R\} \times  \left[\frac{-L}{2}, \frac{-L}{2}\right] \times [\tau_a, \tau_a +\Delta \tau] \right\},
\end{equation}
where $L=4.096 \times 10^2\gg1 $ and $\Delta \tau = 5 \times 10^2\gg 1 $.
In particular, the quantity $\langle \rho^\alpha \rangle$ is obtained numerically via the ``triple-averaging''
\begin{equation}
    \langle \rho^\alpha \rangle = \langle |\psi|^{2\alpha} \rangle = \frac{1}{\Delta \tau \, L} \int_{\tau_a}^{\tau_a + \Delta \tau}  \int_{-L/2}^{L/2} \overline{|\psi(x,t)|^{2\alpha}} \,dx \, dt.
\end{equation}

\subsection{Soliton gas synthesis}

{
Realizations of the bound state soliton gas with the DOS \eqref{linear}
are approximated by $n$-soliton solutions with $n \gg 1$. 
The spatial density of solitons in this gas is given by $\kappa = \int_\Lambda f(\eta) d\eta = 1/4$, hence, since the number of solitons is finite, the approximation is only valid on a spatial interval of length $\ell \sim n / \kappa$, outside which the $n$-soliton solution decays exponentially fast to zero. 

In order to generate the $n$-soliton solution, we use the well-established dressing method originally used within the context of fNLSE soliton gases in \cite{gelash2018strongly}. This method enables solitons to be added recursively to a given solution, one at a time, through a specific algebraic construction. The numerical implementation of this construction is more stable and computationally efficient than solving the Gel'fand-Levitan-Marchenko equations of the inverse scattering transform, thereby enabling the construction of multisoliton wave fields containing a large number of solitons, see the review \cite{suret_soliton_2024} and references therein for a more detailed discussion. 

In this construction the solution is parametrized by the $n$ eigenvalues $\lambda_j = \xi_j+i\eta_j$ of the Zakharov-Shabbat problem in the upper half-plane of $\mathbb{C}$ and $n$ initial associated phases $\theta_j$ and position parameters $x_j^0$. The latter sets of parameters can be gathered into initial norming constants for the dressing method:
\begin{equation}
    C_j(t=0) = \exp(2i \lambda_j x_j^0 + i \theta_j),
\end{equation}
which differs from the norming constants conventionally defined in the inverse scattering transform, see \cite{suret_soliton_2024}.

In order to generate the soliton gas of DOS, we set $\xi_j=0$ (bound state) and distribute the imaginary parts $\eta_j$ with the normalized distribution:
\begin{equation}
    \Phi(\eta) = \frac{f(\eta)}{\kappa} = 2 \eta,
\end{equation}
and the phases $\theta_j$ are uniformly distributed in the interval $[0,2\pi)$. We define 
\begin{equation}
    \kappa_j = \frac{\kappa}{1-\int_\Lambda G(\eta_j,\mu) f(\mu) d\mu} = \frac{2\kappa}{1+\tanh^{-1}(\eta_j) (\eta_j-1/\eta_j)}.
\end{equation}
We observe that, by uniformly distributing the position parameters $x_j^0$ in the intervals $[-n/(2\kappa_j),-n/(2\kappa_j)]$, then the $n$-soliton solution generated by the dressing method evaluated at $t=0$ approximates a realization of the uniform soliton gas with the DOS \eqref{linear} in the spatial region $x\in [-\ell/2,\ell/2]$. We show that the gas is uniform by computing numerically the averages
\begin{equation}
    J_\alpha(x) = \frac{1}{\Delta x} \int_{x-\Delta x/2}^{x+\Delta x/2} \overline{|\psi(x,0)|^{2\alpha}} dx,\quad \alpha=1,2,
\end{equation}
which combine the averages $\overline{|\psi(x,t)|^{2\alpha}}$ defined in \eqref{eq:ens_aver} with the realizations $\psi_k(x,t)$ now being $R$ different $n$-soliton solutions, and the local spatial averaging over a length $\Delta x = 1/\kappa = 4$ which is the average spacing between solitons.
Fig.~\ref{aver_lin} shows that these averages compare well with the corresponding analytical values $\langle \rho^\alpha \rangle$ defined by (see Eq. \eqref{mom}):
\begin{equation}\label{aver_lin}
    \langle \rho \rangle = \int_\Lambda 4\eta f(\eta) d\eta = \frac{2}{3},\quad 
    \langle \rho^2 \rangle = \int_\Lambda \frac{32}{3}\eta^3 f(\eta) d\eta = \frac{16}{15},\quad
\end{equation}
in the spatial region $x\in [-\ell/2,\ell/2]$.}

\begin{figure}
    \centering
    \includegraphics[width=0.6\linewidth]{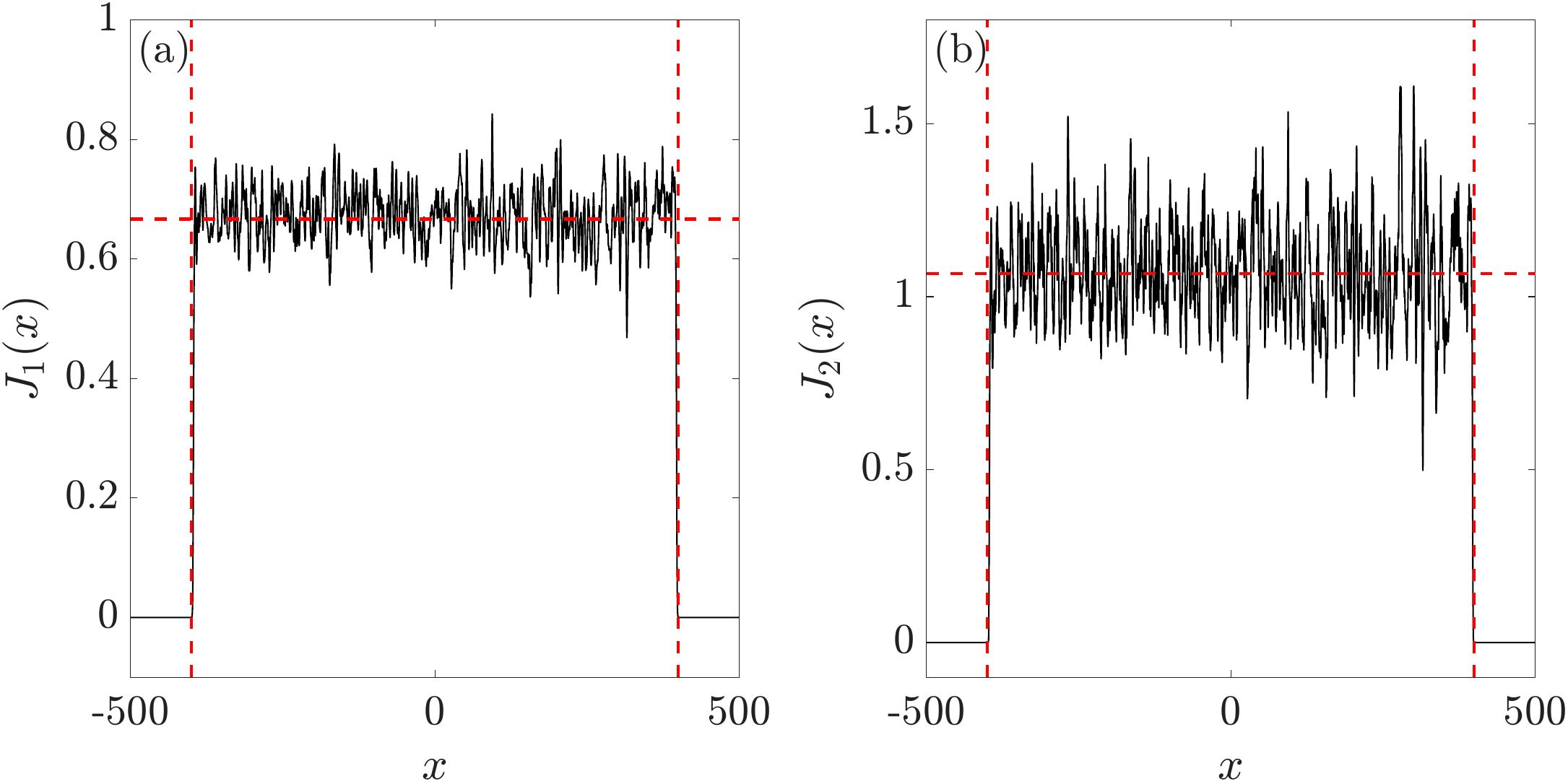}
    \caption{{Averages $J_\alpha(x)$ computed with $R$ different $n$-soliton solutions for the DOS \eqref{linear} where $R=50$ and $n=200$. The vertical dashed lines indicates the position of $x=\pm \ell/2$ and the horizontal dashed lines the values $\langle \rho^\alpha \rangle$ given in \eqref{aver_lin}.}
    }
    \label{fig:aver_lin}
\end{figure}

\end{document}